\documentclass[aps,prl,twocolumn,showpacs,superscriptaddress,longbibliography]{revtex4-2}

\usepackage{graphicx}
\usepackage{color}
\usepackage{amsfonts}

\usepackage{isomath}
\usepackage{amsmath}
\usepackage{amsthm}
\usepackage{amsfonts}
\usepackage{amssymb}
\usepackage{fdsymbol}
\usepackage{braket}
\usepackage{gensymb}

\usepackage[version=4]{mhchem}
\usepackage{physics}

\let\oldAA\AA
\renewcommand{\AA}{\text{\normalfont\oldAA}}

\newcommand{\kbot}[0]{K$_{\text{bot}}$}
\newcommand{\ktop}[0]{K$_{\text{top}}$}

\newcommand{\vect}[1]{\boldsymbol{#1}}

\begin{document}

\title{Observation of dodecagonal replica bands in 30$^{\circ}$-twisted bilayer \ce{WSe_2}}
\author{Christian Valentiner-Branth Fokdal}
\author{Chakradhar Sahoo}
\author{Thomas S. Nielsen}
\author{Alfred J. H. Jones}
\author{Zhihao Jiang}
\affiliation{Department of Physics and Astronomy, Aarhus University, 8000 Aarhus C, Denmark}
\author{Kenji~Watanabe}
\affiliation{Research Center for Electronic and Optical Materials, National Institute for Materials Science, 1-1 Namiki, Tsukuba 305-0044, Japan}
\author{Takashi~Taniguchi}
\affiliation{Research Center for Materials Nanoarchitectonics, National Institute for Materials Science,  1-1 Namiki, Tsukuba 305-0044, Japan}
\author{Marcin Mucha-Kruczyński}
\affiliation{Department of Physics, University of Bath, Bath, United Kingdom}
\author{Søren Ulstrup}
\email{ulstrup@phys.au.dk}
\affiliation{Department of Physics and Astronomy, Aarhus University, 8000 Aarhus C, Denmark}
\date{\today}

\begin{abstract}
Twisted bilayers of two-dimensional (2D) transition metal dichalcogenides are promising systems for achieving tunable quasicrystalline orders with emergent properties.  The underpinning electronic structure and scattering processes in 2D quasicrystals with multiorbital dodecagonal replica bands have so far not been determined.  Here, we utilize angle-resolved photoemission spectroscopy (ARPES) with micrometer spatial resolution to directly observe replica bands in 30$^{\circ}$-twisted bilayer \ce{WSe_2}.  The symmetry and intensity distribution of the observed replicas are explained by interlayer Umklapp scattering from bottom to top  \ce{WSe_2} layers.  Our spectral function measurements are consistent with the presence of a van Hove singularity adjacent to the $\mathrm{K}$-valleys, which underlines the possibility of inducing electronic reconstructions in bilayers with a large interlayer twist angle.
\end{abstract}

\maketitle

Quasicrystals are materials that possess long-range order and symmetries that are forbidden by crystallographic principles in ordinary crystals,  as exemplified by icosahedral Al-Mn alloys and decagonal AlNiCo systems \cite{shechtman1984metallic,levine1984quasicrystals,levine1986quasicrystals,janot_oxford_2012,rotenberg_quasicrystalline_2000,rotenberg_electronic_2004}. 
Several unique topological phases and electron localization effects have been observed in quasicrystalline materials,  which underscore their intermediacy between translation invariant systems and short-range ordered glasses \cite{pierce1993electron,kraus2012topological,else2021quantum}.  Moreover, because quasicrystals represent cross-sections of higher-dimensional crystals, they likewise provide access to quantum Hall effects and fractal patterns beyond three-dimensional space \cite{kraus2013fourd}.

\begin{figure*}
    \centering
    \includegraphics[scale=1]{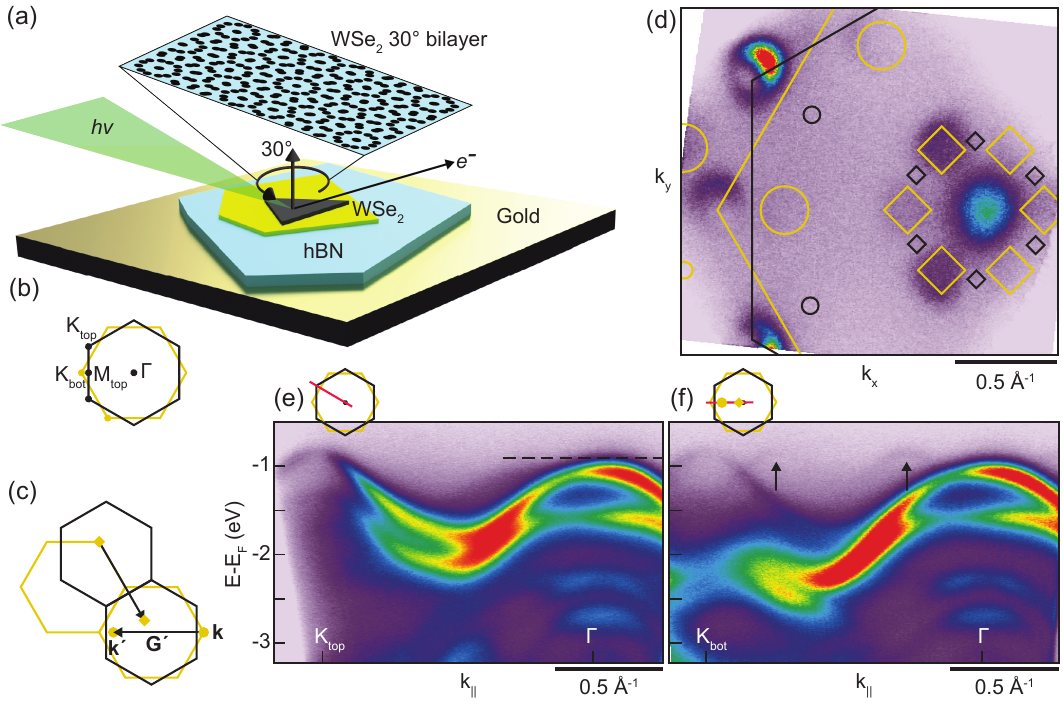}
    \vspace{-0.4 cm}
    \caption{(a) Schematic of micro-focused ARPES experiment on dodecagonal bilayer \ce{WSe_2} supported on hBN on gold-coated \ce{SiO_2}. The blue panel illustrates the 30$^{\circ}$ twist angle-induced structure. The black dots represent the W-atoms.  The incoming photons with energy $h\nu$ are focused on the sample, as illustrated by the green triangle, leading to emission of photoelectrons ($e^-$).   (b) Brillouin zones (BZs) of the 30$^{\circ}$-twisted \ce{WSe_2} flakes.  As in (a), yellow represents the bottom layer and black the top layer.  (c) Overview of the relevant Umklapp scattering processes. (d) ARPES constant energy surface integrated over $(-0.91 \pm 0.05)$ eV. The BZ segments are superimposed in black and yellow lines. Yellow (black) markers denote bottom (top) to top (bottom) layer Umklapp replicas. Circles and diamonds in (c)-(d) demarcate first -and second-order processes, respectively. (e)-(f) ARPES $E(k)$-intensity along (e) $\mathrm{\Gamma}$- \ktop{} and (f) $\mathrm{\Gamma}$- \kbot{}.  The red bars in the BZ insets show the cut directions. The dashed line in (e) indicates the central energy of the isoenergy cut in (d).  Arrows in (f) demarcate replica bands.}
    \label{fig:fig1}
\end{figure*}

Incommensurate twisted bilayer materials can realize quasicrystalline systems on-demand \cite{ahn_dirac_2018, yao_quasicrystalline_2018}.  While small-angle twisted hexagonal bilayers are known for producing large moir\'e supercells and emergent phenomena such as correlated insulators and superconductivity \cite{cao2018correlated,cao2018unconventional,reganMottGeneralizedWigner2020,wang2020correlated,Xia:2024,guoSuperconductivity50degTwisted2025},  a 30$^\circ$ twist between the layers instead leads to a quasicrystal with twelvefold (dodecagonal) symmetry,  as illustrated in Fig.  \ref{fig:fig1}(a). These dodecagonal materials may provide an alternative platform for tuning interlayer couplings, van Hove singularities, correlations and topological properties beyond small-angle moir\'e systems \cite{li2024tuning,Tsang:2024,Xing:2025,Zhida:2026,Chang:2026}. 

Dodecagonal bilayers constructed from twisted bilayer WSe$_2$ are particularly interesting because the valence band of the parent material contains distinct spin,  valley and orbital degrees of freedom,  which undergo spectral reconstructions due to interlayer coupling and Umklapp scattering processes upon forming the quasicrystal \cite{li2024tuning,Zhida:2026}.  The underlying momentum-dependent dispersion of the resulting electronic structure with Umklapp replica bands has not been experimentally determined in the case of 30$^\circ$-twisted bilayer \ce{WSe_2}.  Here,  we present angle-resolved photoemission spectroscopy (ARPES) measurements of such replica bands,  which contain information about the direction of interlayer tunneling between the twisted \ce{WSe_2} layers. Furthermore, we find that the spectral reconstruction around the main and Umklapp $\mathrm{K}$-valleys can be understood in terms of a two-band model with a van Hove singularity at $-1.05$ eV relative to the Fermi level.  

Our dodecagonal quasicrystal was fabricated using mechanical exfoliation of a single layer of \ce{WSe_2},  which was cut and re-stacked as a 30$^\circ$-twisted bilayer on hBN using the tear-and-stack dry-transfer technique \cite{caldwell2010samfab1,castellanos2014deterministic,yi2015samfab2,sahoo2025wse2,suppmat}.  The ARPES measurements were performed at the AU-SGM4 beamline at the ASTRID2 light source \cite{jones2025spatial}.  The light was focused to a 4 $\micro$m spot using a capillary optic. This allowed for singling-out the band structure of the 30$^\circ$-twisted bilayer region formed between the two flakes,  which had dimensions on the order of $10$ $\micro$m \cite{suppmat}.  All ARPES spectra presented in this work were obtained with a photon energy of 60 eV and at room temperature. The experimental energy and momentum resolution were set to 25 meV and 0.02 Å$^{-1}$, respectively.  The experimental setup is summarized in Fig.~\ref{fig:fig1}(a).

The Brillouin zone (BZ) of the bottom (top) WSe$_2$ layer is demarcated by a yellow (black) hexagon in reduced and extended zone schemes in Figs.  \ref{fig:fig1}(b)-(c) and overlaid as a guide on an ARPES constant energy surface extracted at $-0.91$ eV (all energies are given with respect to the Fermi level) in Fig.~\ref{fig:fig1}(d).  ARPES $E(k)$-spectra measured along $\mathrm{\Gamma}$-$\mathrm{K}_{\mathrm{top}}$ and $\mathrm{\Gamma}$-$\mathrm{K}_{\mathrm{bot}}$, where $\mathrm{K}_{\mathrm{bot}}$ ($\mathrm{K}_{\mathrm{top}}$) corresponds to a $\mathrm{K}$-point in the bottom (top) WSe$_2$ layer, are shown in Figs.~\ref{fig:fig1}(e)-(f).  The ARPES spectra reveal an intense valence band with a twofold splitting of 0.51 eV at $\mathrm{\Gamma}$,  signifying interlayer coupling between the two WSe$_2$ layers \cite{Jones:2022},  as well as 0.47 eV spin-orbit split bands around $\mathrm{K}$-valleys at the corners of the two $30^{\circ}$-rotated BZs.  The bands around the $\mathrm{K}$-valleys are composed of in-plane orbitals, resulting in minimal interlayer coupling around $\mathrm{K}$, as opposed to the $\mathrm{\Gamma}$-valley bands that are composed of out-of-plane orbitals \cite{kormanyos2015k}.  The layer-localized bands around $\mathrm{K}_{\mathrm{bot}}$ thus exhibit weaker ARPES intensity compared to $\mathrm{K}_{\mathrm{top}}$ due to the short inelastic mean free path of the photoelectrons \cite{moser2017experimentalist}.  

Strikingly,  in Fig.  \ref{fig:fig1}(f) the ARPES spectrum reveals a set of new features (see arrows) between the main $\mathrm{K}_{\mathrm{bot}}$- and $\mathrm{\Gamma}$-valley bands.  These and several others that are observed throughout the measured segment of the BZs are marked by yellow diamonds and circles in the constant energy surface in Fig.  \ref{fig:fig1}(d).  The dispersion of these features matches the dispersion of the main  $\mathrm{K}$-valley bands, but with the intensity of the lower spin-orbit split branch masked by the intensity of main bands,  thus giving a strong clue to their origin as $\mathrm{K}$-valley replica bands. 

We interpret the replicas that contribute to the ARPES intensity as a result of interlayer tunneling from bottom to top WSe$_2$ layer,  which can be understood in terms of Umklapp scattering \cite{bistritzer2011moire,koshino2015interlayer}: The bottom (top) layer state  $\ket{\vb{k}+\vb{G}}$, where $\vb{k}$ is the crystal momentum and $\vb{G}$ is a reciprocal lattice vector of the bottom (top) layer, scatters into the top (bottom) layer state $\ket{\vb{k'}}$ using a reciprocal lattice vector of the top (bottom) layer, $\vb{G}'$, and the condition of momentum conservation within a reciprocal lattice vector, $\vb{k}+\vb{G} = \vb{k'}+\vb{G}'$. The final state at $\vb{k'}$ contains information about the initial state and is referred to as an Umklapp replica.  Two examples of Umklapp scattering that lead to observable ARPES replicas are shown in Fig.~\ref{fig:fig1}(c), where the black arrows are the $\vb{G}$ vectors of the top layer and the yellow markers are the replicas.  We refer to scattering within the first BZ as first-order replicas and scattering from the second BZ as second-order replicas,  which have been demarcated in Figs.  \ref{fig:fig1}(c)-(d) by circles and diamonds,  respectively.  The presence of both first- and second-order replicas,  the distances between replicas and main bands in $k$-space and the alignment of BZs extracted from Fig.~\ref{fig:fig1}(d) altogether ascertain that a twist angle of $30^{\circ}$ has been achieved.  Interestingly,  we do not observe intensity from first (second) order replicas in regions where bands originating in the top layer scatter into the bottom layer,  which would be expected to occur around the black circles (diamonds) in Fig.~\ref{fig:fig1}(d). The replica intensity in dodecagonal bilayer \ce{WSe_2} is thus characterized by a distinct layer asymmetry. 

\begin{figure}
    \centering
    \includegraphics[scale=1]{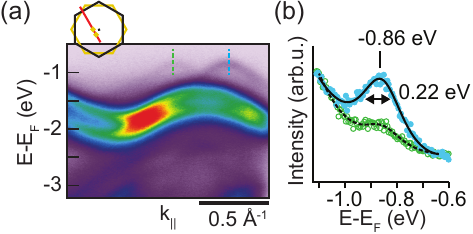}
    \vspace{-0.4 cm}
    \caption{(a) ARPES spectrum along a cut containing two second-order replicas.  The red bar in the BZ inset marks the cut direction.  (b) EDCs extracted along the green and blue dotted lines in (a), represented with open and filled circles, respectively. The full and dashed curves overlaid on the data represent fits to a Voigt peak on a polynomial background.  The peak is centred at -0.86 eV and has Lorentzian full width at half maximum (FWHM) of 0.22 eV,  as illustrated by double-headed arrows.}
    \label{fig:cuts}
\end{figure}

The intensity of the six bottom-layer second-order replicas observed around $\mathrm{\Gamma}$ in Fig.~\ref{fig:fig1}(d) displays threefold rather than sixfold symmetry.  This is further visualized in Fig.  \ref{fig:cuts}(a)-(b) where an $E(k)$-ARPES cut crossing two second-order replica bands and energy distribution curves (EDCs) through their band maxima are presented. EDC fits reveal that the replicas exhibit a band maximum at -0.86 eV,  which coincide with the energy of the main valence band maxima at  $\mathrm{K}_{\mathrm{bot}}$ and $\mathrm{K}_{\mathrm{top}}$. The linewidth of both replica bands is 0.22 eV, as measured in terms of the full width at half maximum (FWHM) of the EDC fit. The intensity differs by a factor of 3 between the two bands.  This intensity modulation is a consequence of interference between the emitted photoelectrons, caused by the threefold symmetry of the individual layers \cite{Rostami:2019}. Additionally, the intensity of all second-order replicas to the right of the $\mathrm{\Gamma}$-point in Fig.~\ref{fig:fig1}(d) is further reduced,  which is attributed to odd terms in the photoemission dipole matrix element involving the WSe$_2$ states on the other side of the mirror line crossing $\mathrm{\Gamma}$ \cite{moser2017experimentalist}.

In the following, we develop an expression for the ARPES spectral function of the replicas that provides more insight into the observed layer asymmetry of intensity patterns. For simplicity, we assume that only states from the top valence band of the top and bottom layer mix and limit the scattering to first-order processes. Then, a state $\ket{\vect{k},\mathrm{b}}$ in the bottom (b) layer mixes with states $\ket{\vect{k}',\mathrm{t}}$ in the top (t) layer to produce a state,
\begin{align}\label{eqn:bottom_layer}
|\vect{k},\mathrm{b}) = c^{\mathrm{b}}_{|\vect{k}|}\ket{\vect{k},\mathrm{b}} + \sum_{\vect{G}_{\mathrm{bot}}-\vect{G}_{\mathrm{top}}} c^{\mathrm{t}}_{|\vect{k}+\vect{G}_{\mathrm{bot}}|} \ket{\vect{k}+\vect{G}_{\mathrm{bot}}-\vect{G}_{\mathrm{top}},\mathrm{t}}. 
\end{align} 
Above, $|...)$ denotes the mixed state and $\ket{...}$ an unperturbed state of one of the layers. The complex numbers $c^{l}_{|\vect{k}|}$,  with $l=\{\mathrm{b},\mathrm{t}\}$, are the expansion coefficients. Additionally, in the spirit of continuum models originally constructed for twisted bilayer graphene, we assume that in the momentum space the interlayer coupling is isotropic \cite{bistritzer2011moire,koshino2015interlayer,thompson2020determination}. In the absence of interlayer coupling, $c^{\mathrm{b}}_{|\vect{k}|}=1$ and $|\vect{k},\mathrm{b}) =\ket{\vect{k},\mathrm{b}}$. Nonzero interlayer coupling leads to mixing of the states of the two layers and hence nonzero $c^{\mathrm{t}}_{|\vect{k}+\vect{G}_{\mathrm{bot}}|}$ for states that started as pure bottom layer states. Similarly, for the top layer, one can write,
\begin{align*}
|\vect{k},\mathrm{t}) = c^{\mathrm{t}}_{|\vect{k}|}\ket{\vect{k},\mathrm{t}} + \sum_{\vect{G}_{\mathrm{top}}-\vect{G}_{\mathrm{bot}}} c^{\mathrm{b}}_{|\vect{k}+\vect{G}_{\mathrm{top}}|} \ket{\vect{k}+\vect{G}_{\mathrm{top}}-\vect{G}_{\mathrm{bot}},\mathrm{b}}. 
\end{align*} 

Within this approximation, the single-particle spectral function measured by ARPES \cite{Damascelli:2004}, $A(\vect{k}_{\mathrm{replica}},E)$, at the replica wave vector $\vect{k}_{\mathrm{replica}}$, is, 
\begin{align}\begin{split}
&A(\vect{k}_{\mathrm{replica}},E)= 
\sum_{l,l'=\mathrm{t},\mathrm{b}}|\bra{\vect{k}_{\mathrm{replica}},l}\vect{k},l')|^{2}\delta(E-E_{\vect{k}}^{l'}),
\end{split}\end{align}

where $E_{\vect{k}}^{l'}$ is the band dispersion of layer $l'$.  For replicas in incommensurate twisted bilayers, we have $\vect{k}_{\mathrm{replica}}\neq\vect{k}$ and $\vect{G}_{\mathrm{bot}}-\vect{G}_{\mathrm{top}}\neq0$. Moreover, each replica (i) fulfills momentum conservation modulo a reciprocal lattice vector for a different set of $\vect{G}_{\mathrm{bot}}$ and $\vect{G}_{\mathrm{top}}$ and (ii) can be traced back to the original unperturbed state in either the top or bottom layer. The former means that only one term is selected from the sum over reciprocal vectors $\vect{G}_{\mathrm{bot}}-\vect{G}_{\mathrm{top}}$ in \eqref{eqn:bottom_layer}. The latter, in turn, implies that only the contribution with either $l'=\mathrm{b}$ or $l'=\mathrm{t}$ is relevant at energy $E$ at a specific $\vect{k}_{\mathrm{replica}}$. Altogether, for a  replica of a state of the bottom layer, $l'=\mathrm{b}$, originally at wave vector $\vect{k}$, 
\begin{align}
A(\vect{k}_{\mathrm{replica}}\neq\vect{k},E) & \propto \delta_{\vect{k}_{\mathrm{replica}},\vect{k}+\vect{G}_{\mathrm{bot}}-\vect{G}_{\mathrm{top}}} \left| c_{|\vect{k}+\vect{G}_{\mathrm{bot}}|}^{\mathrm{t}} \right|^{2}.
\end{align}
The ARPES spectral function thus provides a measure of how much of the state with wave vector $\vect{k}_{\mathrm{replica}}$ in the layer into which the electron scattered is mixed with the original state at $\vect{k}$. In other words, it represents the amount of spectral weight moved from the original state to the mixed one.  In the interlayer coupling process this spectral weight moves from one layer to the other. Because ARPES is depth sensitive, the same amount of spectral weight that moved from the bottom layer to the top one results in a higher intensity feature than if it moved from the top layer to the bottom one.  

The above analysis supports the initial interpretation that the observed replicas are a result of spectral reconstruction by the interlayer coupling (initial-state effect) rather than final-state effects. The latter would involve photoelectron scattering on any linear combination of the available reciprocal vectors as it exits the material \cite{polley2019finalstate}, meaning both top and bottom replicas would be visible.  Moreover, our experiments are performed at room temperature and we expect the secondary scattering effects to be weak due to the temperature dependence of the Debye-Waller factor for the diffracted photoelectrons \cite{bostwick2007symmetry,peng2005electron}. Finally, our interpretation suggests that replicas of the $\mathrm{\Gamma}$-point from the second BZ of the bottom layer should also be visible in our data, which they are as we show in the Supplemental Material \cite{suppmat}.

\begin{figure}
    \centering
    \includegraphics[scale=1]{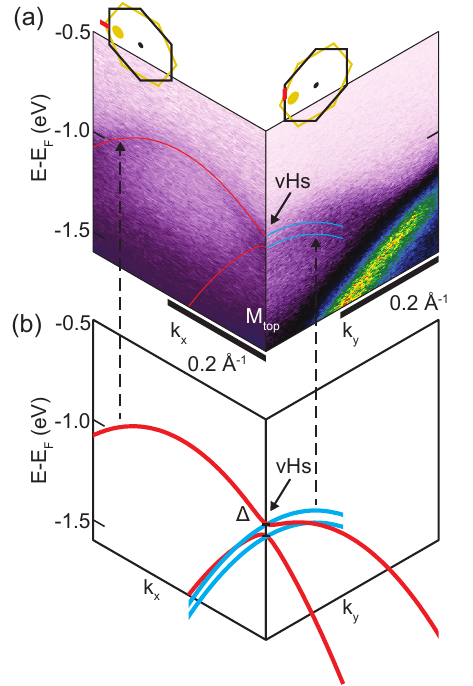}
    \vspace{-0.4 cm}
    \caption{(a) ARPES $E(k)$-dispersion cuts around the crossing between the main band of the bottom layer and a first order replica extracted along $\mathrm{\Gamma}$-\kbot{} (left panel) and \ktop{} -M$_{\text{top}}$ (right panel). The red bars in the BZ insets show the cut directions. (b) Extracted dispersion around the band crossing based on fitting a gapped two-band model to the ARPES spectra in (a) \cite{suppmat}.  The red (blue) curves correspond to the bands along $k_x$ ($k_y$).  The arrow points to a van Hove singularity (vHs) at -1.05 eV where the band curvature changes sign between the two orthogonal cuts.  Dashed arrows indicate the band positions in the ARPES spectra.  An assumed gap of $\Delta=0.06$ eV at the band crossing is marked by a bracket.}
    \label{fig:fig3}
\end{figure}

As a final point, we address the $E(k)$-dispersion around the crossing at the $\mathrm{M}_{\text{top}}$-point between replica and main bands, as shown in Fig.  \ref{fig:fig3}(a).  This is the only such crossing we are able to examine in detail as in all other cases the crossings are masked by nearby intense main bands,  for example those seen in Fig.  \ref{fig:cuts}(a).  The spectra in Fig.  \ref{fig:fig3}(a) correspond to two orthogonal cuts through M$_{\text{top}}$.  By using a gapped two-band model to describe hybridization at the crossing, we simulate the measured ARPES intensity and extract the dispersion in the region around $\mathrm{M}_{\text{top}}$, as discussed in detail in the Supplemental Material \cite{suppmat}.  In the model, we assume a gap of 0.06 eV,  which is the minigap determined by scanning tunneling spectroscopy measurements of 30$^\circ$- twisted bilayer WSe$_2$ \cite{li2024tuning}.  From the simulations, it is clear that such a gap cannot be directly resolved in the ARPES spectrum because it is substantially smaller than the linewidth of the replicas determined to be 0.22 eV in Fig.  \ref{fig:cuts}(b),  although indications of a gap are present via an elongation of the intensity at the center of the band crossing, as discussed in the Supplemental Material \cite{suppmat}.  

The dispersion extracted from the fit of the two-band model to the ARPES spectra is shown in Fig. \ref{fig:fig3}(b).  At $\mathrm{M}_{\text{top}}$ at an energy of -1.05 eV,  the top-most band has positive curvature along $k_x$ and negative curvature along $k_y$.  Hence,  a saddle point occurs at -1.05 eV, which will lead to a van Hove singularity (vHs) in the density of states.  The vHs occurs within 0.2 eV of the valence band maximum at $\mathrm{K}$ and in the middle of the spin-split branches dispersing around the $\mathrm{K}$-valley.  Van Hove singularities are often associated with observations of instabilities and strong correlations.  Thus,  the presence of a vHs near the $\mathrm{K}$-valleys makes 30$^\circ$-twisted bilayer WSe$_2$ an interesting candidate for inducing emergent phenomena in states with distinct spin and orbital characters.  In order to bring the vHs within the regime of electron transport it is necessary to induce strong hole doping in twisted bilayer WSe$_2$,  which is possible using  RuCl$_3$ and NbSe$_2$ substrates \cite{packChargetransferContactsMeasurement2024,nielsen:2026,Clark:2026}.  One would also expect new optical transitions with strongly enhanced exciton populations for optical excitations that are resonant with the vHs energy.  

In summary, we directly observe dodecagonal replica bands in quasicrystalline 30$^\circ$-twisted bilayer WSe$_2$. The measured ARPES spectral function can be understood in terms of Umklapp scattering from bottom- to top-layer $\mathrm{K}$-valleys and are well described by a gapped two-band model with a saddle point.  Our findings highlight large-angle twisted bilayer transition metal dichalcogenides as an interesting platform for studying transport and optoelectronic properties.

\section{Acknowledgements}
The work was funded/co-funded by the European Union (ERC grant EXCITE with project number 101124619). Views and opinions expressed are however those of the author(s) only and do not necessarily reflect those of the European Union or the European Research Council. Neither the European Union nor the granting authority can be held responsible for them. The authors acknowledge funding from the Novo Nordisk Foundation (Project Grant NNF22OC0079960) and VILLUM FONDEN under the Villum Ascending Investigator Program (VIL83457). C.S. acknowledges Marie Sklodowska-Curie Postdoctoral Fellowship (project 101059528 MaPWave). K.W. and T.T. acknowledge support from the JSPS KAKENHI (Grant Numbers 21H05233 and 23H02052), the CREST (JPMJCR24A5), JST and World Premier International Research Center Initiative (WPI), MEXT, Japan.

\newpage

\begin{widetext}
\section{Supplementary Information}

\vspace{1cm}

\section{Sample fabrication}

The sample was fabricated using mechanical exfoliation followed by a dry-transfer stacking technique \cite{caldwell2010samfab1,castellanos2014deterministic,yi2015samfab2,sahoo2025wse2}. Monolayer \ce{WSe_2} flakes were exfoliated from a bulk \ce{WSe_2} crystal onto polydimethylsiloxane (PDMS) Gel-Pak (PF-40×40-0170-X4) substrates using Scotch/Nitto tape and identified by optical microscopy. A selected monolayer flake was transferred onto an Si/\ce{SiO_2} substrate which was prepared prior to the transfer by oxygen plasma cleaning for 6 minutes. Subsequently, the flake was cleaved into two pieces using a sharp needle with a $\sim$2 $\micro$m tip. The twisted bilayer (TB) was assembled using a polycarbonate (PC) film supported on a PDMS stamp. The two \ce{WSe_2} flakes, originating from the same parent crystal, were sequentially picked up and stacked with a relative twist angle of 30° in a micromanipulator, ensuring precise control of their crystallographic orientation. The bilayer was then transferred onto an exfoliated hexagonal boron nitride (hBN) flake ($\sim$30 nm thick). The TB on hBN stack was transferred onto a Si/\ce{SiO_2} substrate with a 50nm Au layer on top. Part of the \ce{WSe_2} is in direct contact with the Au layer to provide electrical grounding during the photoemission experiment. After assembly, the sample was cleaned by sequential immersion in chloroform, acetone, and isopropanol for 1 min each to remove residual PC. Finally, the sample was annealed at 250 °C for 5 h in forming gas (5\% \ce{H_2} in Ar), further reducing polymer residues and improving the interfacial adhesion of the stacked layers. Additional details of the exfoliation and dry-transfer procedures are described in reference \cite{sahoo2025wse2}. 

An optical microscopy image of the final sample structure, is shown in Fig.  \ref{fig:samplefig}. The substrate (gold), hBN flake, and \ce{WSe_2} flake, which contains monolayer (ML), bilayer (BL) and multilayer (marked \ce{WSe_2}) regions, as well as the the TB (outlined in white), are clearly seen. The scale bar indicates the size of the relevant regions, showing that the TB has dimensions on the order of 10 $\micro$m. 

Prior to measurements the sample plate was annealed overnight at 200 $^{\circ}$C in the load-lock of the end-station, where the pressure is below $5\cdot10^{-8}$ mbar, to remove water and other potential residues.

\begin{figure*}[h]
    \centering
    \includegraphics[scale=1]{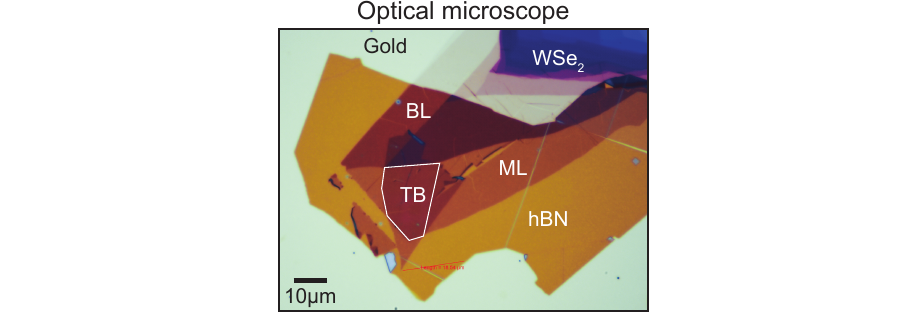}
    \vspace{-0.4cm}
    \caption{Optical microscope image of the sample. The regions of monolayer (ML), bilayer (BL), twisted bilayer (TB), hBN, and gold are indicated. The blue region near the top is multilayer \ce{WSe_2} acting as electrical contact to the gold.}
    \label{fig:samplefig}
\end{figure*}

\newpage

\section{Umklapp replica of $\mathrm{\Gamma}$-band originating in bottom layer}

Figure \ref{fig:gammarep}(a) shows the expected scattering path for a bottom-to-top Umklapp replica from the second BZ of the bottom layer.  Figure \ref{fig:gammarep}(b) shows the observed replica at this position, cresting the top-layer $\mathrm{K}$-band. We do not observe any similar bands at the positions where one would expect a top-to-bottom replica, illustrated by Figs. \ref{fig:gammarep}(c)-(d).  Within the measured segment of $k$-space, another $\mathrm{\Gamma}$-replica from the bottom layer should be present, but remains unseen. This, however, is consistent with suppression from the threefold symmetric intensity modulation present in the K-valley replicas, which is sufficient to submerge the missing $\mathrm{\Gamma}$-replica in the background.

\begin{figure}[h]
    \centering
    \includegraphics[scale=1]{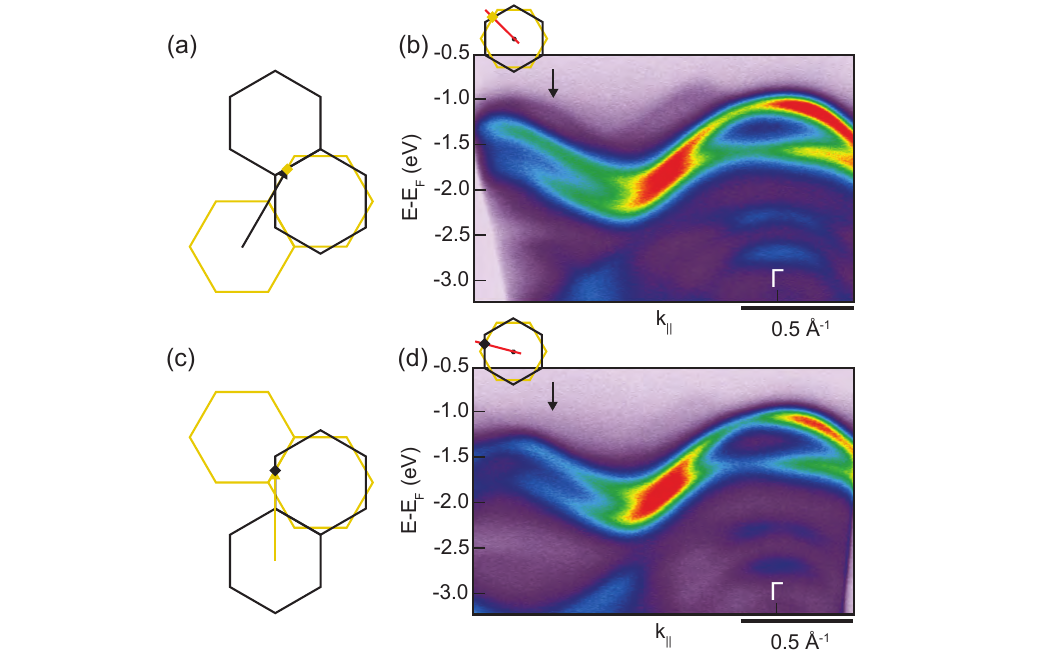}
    \vspace{-0.4cm}
    \caption{(a) Scattering path for a bottom-to-top $\mathrm{\Gamma}$-replica band using a top-layer reciprocal lattice vector (black arrow).  (b) ARPES intensity measured along the cut shown in the BZ inset,  which contains the $\mathrm{\Gamma}$-replica originating from the process shown in (a).  The black arrow demarcates the replica. (c) Scattering path for a top-to-bottom $\mathrm{\Gamma}$-replica band, using a bottom-layer reciprocal vector (yellow arrow). (d) ARPES intensity along the direction shown in the BZ inset, through the expected position of a top-to-bottom $\mathrm{\Gamma}$-replica. The arrow indicates where the missing replica should be.}
    \label{fig:gammarep}
\end{figure}

\newpage

\section{ARPES intensity simulations within two-band model}
We construct a simple two-band model that we use to simulate the measured ARPES intensity in order to extract the band dispersion and explore the implications of spectral reconstructions introduced by band hybridization between $\mathrm{K}$-valley main and replica bands.  In particular,  the simulations outline how the magnitude of the hybridization-induced gap and the intrinsic linewidth of bands affect the energy distribution curves (EDCs).

The following analysis is based on the band crossing around $\mathrm{M}_{\mathrm{top}}$,  which is discussed in Fig. 3 of the main manuscript. Figure \ref{fig:S3} presents the same data with the full crossing displayed along both $k_x$ and $k_y$ with and without overlaid bands, applied in the simulation, for clarity.  Note that the spectrum looks broader along $k_y$ because the dispersion is composed of two closely spaced bands with the same curvature in this direction (see blue overlaid bands in Fig.  \ref{fig:S3}).

\begin{figure*}[h]
    \centering
    \includegraphics[scale=0.79]{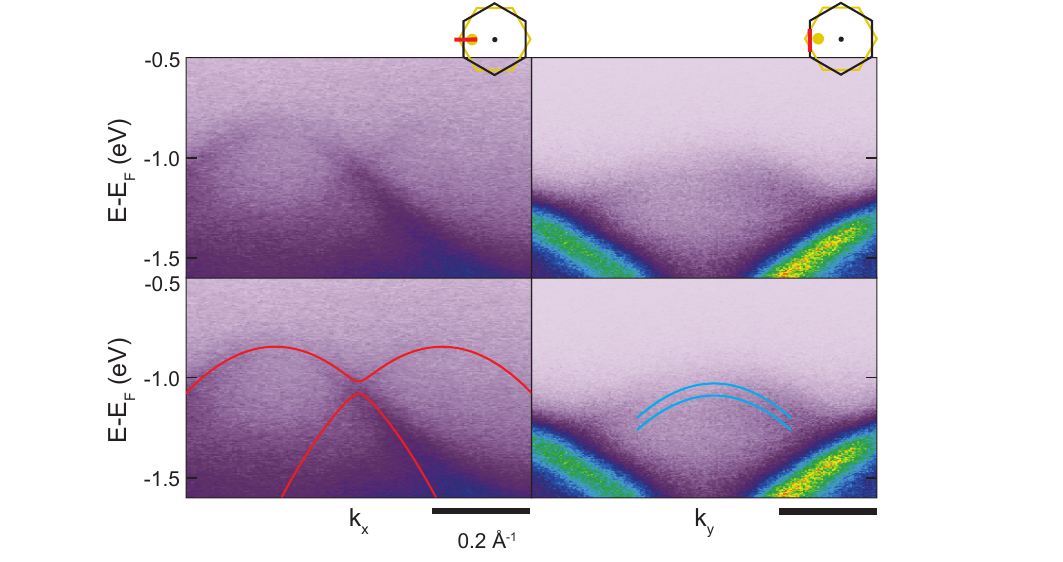}
    \vspace{-0.4cm}
    \caption{ARPES intensity around $\mathrm{M}_{\mathrm{top}}$ along $k_x$ and $k_y$ with (bottom panels) and without (top panels) overlaid bands extracted from fits to simulations of the ARPES intensity.  The red bars in the BZ insets illustrate the cut directions.  The same data is presented in Fig.  3 of the main manuscript. }
    \label{fig:S3}
\end{figure*}

We utilize the cut along the $k_x$-direction in Fig.  \ref{fig:S3} to optimize the simulated ARPES intensity in comparison to the data.  The ARPES intensity, $I(\vb{k},E)$,  is described as:

\begin{align}
    I(\vb{k},E) \propto \sum_{f,i}|M^{\vb{k}}_{f,i}|^2 A(\vb{k},E)f(E)*G(\vb{k},E),
\end{align}
where $\vb{k}$ is the crystal momentum, $E$ is the energy and $f(E)$ is the Fermi-Dirac distribution. The single-electron matrix elements $|M^{\vb{k}}_{f,i}|^2$ connect the initial ($i$) and final ($f$) states, and depend on both $\vb{k}$ and $E$. Finally, $A(\vb{k},E)$ is the spectral function describing the unperturbed single-particle dispersion $\epsilon_{\vb{k}}$, dressed by the many-particle interactions encoded in the electronic self-energy $\Sigma=\Sigma'+i\Sigma''$. Furthermore, the experimental $E$ and $k$ resolution is included via a convolution with a Gaussian distribution. The spectral function is written as:
\begin{align}
    A(\vb{k},E)&=\frac{\Sigma''}{(E-\epsilon_{\vb{k}})^2+\Sigma''^2},
\end{align}
where we set $\Sigma' = 0$ for simplicity,  and $\epsilon_{\vb{k}}$ is described in terms of a gapped two-band dispersion with upper ($+$) and lower ($-$) bands given by

\begin{align}
E_{\pm}&=\frac{1}{2}\qty[E_1+E_2\pm\sqrt{(E_1-E_2)^2+\Delta^2}].
\end{align}
The unperturbed contributions are described as parabolic in $k$:
\begin{align}
E_i(k)&=C+a(k-k_{0,i})^2, 
\end{align}
where $i=\left\{1,2\right\}$,  $\Delta$ is the gap size, $C$ is the band maximum,  $a$ the curvature and $k_{0,i}$ the band position in momentum.  The intensity variation within each band is modelled through the $E$- and $k$-dependent matrix elements.  The background intensity is adjusted to match the experimental background.  As a starting point we set $\Sigma''=0.11$ eV to be consistent with the full width at half maximum (FWHM) of 0.22 eV determined in Fig. 2(b) of the main paper. 

A close-up of the measured band crossing is presented in the top-left panel of Fig.  \ref{fig:simgaps}.  The crossing resembles an elongated region of intensity,  which we find is best described by the simulation if we include a gap given by $\Delta=0.06$ eV,  as measured in scanning tunneling microscopy experiments on 30$^\circ$  twisted bilayer WSe$_2$ \cite{li2024tuning}.  A fit of the simulated intensity to the data can be evaluated using the $\chi^2$-metric.  For the simulated intensity with and without gaps respectively, this results in values of $\chi^2=156.6$ and $\chi^2=179.2$, meaning there is a slightly better agreement for the model with a gap.  The elongation or "neck" effect of having a gap between the two bands is illustrated using simulations with increasing gap in Fig.  \ref{fig:simgaps}.  The corresponding EDC through the band crossing develops from a peak to a plateau and into a shoulder for a gap larger than 0.09 eV.  In Fig.  \ref{fig:simwidths} we simulate the effect on the spectra by reducing the linewidth as specified by $\Sigma''$ for a fixed gap of 0.06 eV.  When  $\Sigma''$ is reduced below 0.08 eV a noticeable shoulder emerges in the EDC through the band crossing.  To summarize,  the results in Figs.  \ref{fig:simgaps}-\ref{fig:simwidths} outline the significance of gap size and linewidth on the ARPES $(E,k)$-intensity of hybridized bands around the crossing, and how these factors can obfuscate hydridisation gaps.

\begin{figure*} [h]
    \centering
    \includegraphics[scale=0.89]{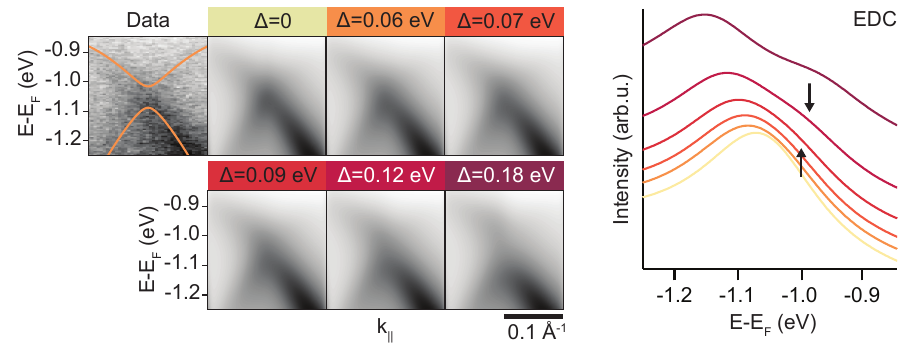}
     \vspace{-0.4cm}
    \caption{Simulation of ARPES intensity around band crossing within two-band model with varying gap size.  The top left panel displays measured data with optimum bare bands overlaid.  The remaining panels to the left present simulated ARPES spectra with the stated gap sizes.  The right panel presents EDCs through the centre of the gap.  The colour of the curves is matched to the color of the box above the simulated spectra with the corresponding gap size.  Arrows indicate a shoulder in the EDCs caused by the gap.}
    \label{fig:simgaps}
\end{figure*}

\begin{figure*} [h]
    \centering
    \includegraphics[scale=0.89]{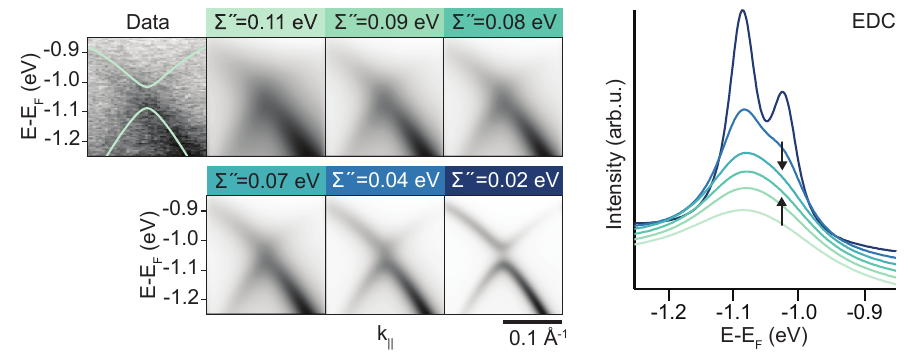}
     \vspace{-0.4cm}
    \caption{Simulation of ARPES intensity around band crossing within two-band model with varying linewidth set by $\Sigma''$.  The top left panel displays measured data with optimum bare bands overlaid.  The remaining panels to the left present simulated ARPES spectra with the stated values of $\Sigma''$.  The right panel presents EDCs through the centre of the gap.  The colour of the curves is matched to the color of the box above the simulated spectra with the corresponding gap size.  Arrows indicate a shoulder in the EDCs caused by the gap.}
    \label{fig:simwidths}
\end{figure*}

\newpage
\end{widetext}

\end{document}